\documentclass[aps,prl,twocolumn,showpacs,superscriptaddress]{revtex4-2}
\usepackage{amsmath,amssymb}
\usepackage{graphicx}
\usepackage{wasysym}
\usepackage{amsfonts}
\usepackage{bm}
\usepackage{enumerate}
\usepackage{color}
\usepackage{url}
\usepackage{hyperref}
\usepackage{epstopdf}
\usepackage{latexsym}
\usepackage[caption=false,singlelinecheck=false]{subfig}

\newcommand{\bea}{\begin{eqnarray}}
\newcommand{\eea}{\end{eqnarray}}
\newcommand{\bs}{\boldsymbol}

\newcommand{\Grad}{\bs \nabla}
\newcommand{\D}{{\bs D}}

\begin{document}
\title{Pair Density Waves and Supercurrent Diode Effect in Altermagnets}
	
\author{GiBaik Sim}
\affiliation{Department of Physics and Astronomy, The University of Tennessee, Knoxville, Tennessee 37996, USA}
\affiliation{Materials Science and Technology Division, Oak Ridge National Laboratory, Oak Ridge, Tennessee 37831, USA}
\affiliation{Department of Physics, Hanyang University, Seoul 04763, Republic of Korea}
	
\author{Johannes Knolle}
\affiliation {Department of Physics TQM, Technische Universit\"{a}t M\"{u}nchen, $\&$ James-Franck-Straße 1, D-85748 Garching, Germany}
\affiliation{Munich Center for Quantum Science and Technology (MCQST), 80799 Munich, Germany}
\affiliation{Blackett Laboratory, Imperial College London, London SW7 2AZ, United Kingdom}
	
\date{\today}

\begin{abstract}
Metallic altermagnets are unusual collinear magnets that feature zero net magnetization with momentum-dependent spin splitting. Here, we show that this spin splitting can induce pair density wave states even in the absence of external magnetic fields. Focusing on BCS-type attractive interactions, we find the stabilization of symmetrically distinct pair density wave states depending on the chemical potential. These states include Fulde-Ferrell and Fulde-Ferrell* states, both of which break inversion symmetry. We investigate the supercurrent properties and discover non-reciprocal supercurrents for both the Fulde-Ferrell and Fulde-Ferrell* states with distinct spatial dependencies. We propose that the supercurrent diode effect can serve as an experimental tool for distinguishing between different pair density waves in metallic altermagnets and discuss the relation to material candidates.
\end{abstract}

\maketitle
\textbf{\textit{Introduction.---}}
Collinear magnets are commonly thought to display either ferromagnetic or antiferromagnetic order~\cite{white1983quantum,fazekas1999lecture}. Ferromagnets (FM) break time-reversal symmetry with a net magnetic moment leading to spin split electronic bands. On the other hand, traditional antiferromagnets (AFM) show zero net magnetization and are symmetric under the time reversal, which flips the spin, followed by a lattice translation, resulting in Kramer's spin degenerate bands in momentum space. However, recent investigations have challenged this binary classification, proposing a new subtype of collinear magnetism termed altermagnets (AM), which are characterized by momentum-dependent spin split bands with zero net magnetization~\cite{yuan2020giant,vsmejkal2020crystal,vsmejkal2022beyond,vsmejkal2022emerging,feng2022anomalous,steward2023dynamic}. In AM the symmetry that connects the magnetic sublattices  with opposite spins are non-trivial rotations ~\cite{PhysRevB.109.024404}, for example a fourfold rotation leads to $d$-wave AM on the square lattice~\cite{leeb2024spontaneous,das2024realizing,rao2024tunable}.
	
The interplay between magnetism and superconductivity can lead to qualitatively new  phenomena, for example a change of pairing symmetry~\cite{bergeret2005odd,scalapino2012common}, and has been a central focus in condensed matter physics research~\cite{dai2012magnetism,fradkin2015colloquium}. The the spin splitting of electronic bands in FM impedes the formation of conventional $s$-wave spin-singlet superconductivity as the formation of Cooper pairs is energetically suppressed. However, an alternative is to stabilize Cooper pairs with  a finite center of mass momentum, as proposed to be realizeable in FM/superconductor junctions~\cite{demler1997superconducting,ryazanov2001coupling,kontos2002josephson}. Here, we study the effect of the unique momentum-dependent spin splitting of AM on superconductivity. Theoretical investigations have focused on cases where superconductivity is induced solely through the proximity effect within AM/superconductor junctions~\cite{ouassou2023dc,beenakker2023phase,PhysRevB.108.L060508,zhang2024finite,wei2024gapless,giil2024superconductor,sun2023andreev,ghorashi2024altermagnetic,li2023majorana,maeda2024theory,lu2024varphi}. 
Nevertheless, the understanding of bulk superconductivity in AM has been limited~\cite{brekke2023two,maeland2024many,PhysRevB.108.184505,sumita2023fulde,chakraborty2024zero,bose2024altermagnetism,hong2025unconventional}, despite the presence of superconductivity in AM materials, which include RuO$_2$ thin films with strain~\cite{uchida2020superconductivity,ruf2021strain} and possibly hole-doped La$_2$CuO$_4$~\cite{vsmejkal2022beyond,dai2012magnetism}.
\begin{figure}[]
\includegraphics[width = 1.0\columnwidth]{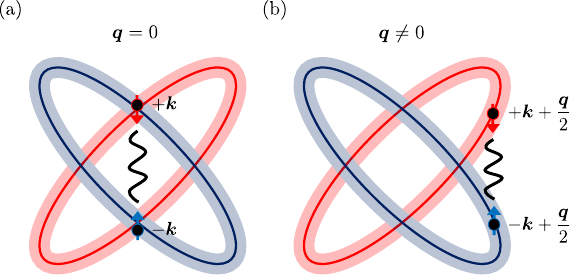}		
\caption{(color online) Two distinct spin-polarized Fermi surfaces (red and blue ellipses) and the regions nearby in energy within the Debye frequency $\omega_D$. (a) For conventional spin-singlet $\bs q\!=\!0$ state, the pairing only happens where two spin states overlap. (b) For spin-singlet $\bs q\!\neq\!0$ states, the pairing can also happen where the spin split bands do not overlap. }
\label{fig_1}
\end{figure}

In this work, we explore the bulk superconductivity arising from a conventional BCS-type attractive interaction between electrons in $d$-wave AM metals. We find the emergence of pair density wave (PDW) states even in the absence of an external magnetic field or charge density wave instabilities. Deriving the Ginzburg–Landau theory from a microscopic model, we find the stabilization of inversion symmetry-breaking PDW states like Fulde–Ferrell (FF) and Fulde–Ferrell* (FF*) states. Further investigation into supercurrent properties reveals non-reciprocal supercurrents in the FF state along the $x$ direction and in the FF* state along $x$ and $y$ directions.
In a related context, a recent work studied the symmetry properties of possible supercurrent diode effects, e.g. non-reciprocal critical supercurrents~\cite{yuan2022supercurrent,he2022phenomenological,daido2022intrinsic,scammell2022theory,lu2023tunable,cayao2024enhancing}, in altermagnetic superconductors and related  heterostructures~\cite{banerjee2024altermagnetic}. Here we derive the effective Ginzburg-Landau theory and show that non-reciprocal supercurrents may appear for intrinsic inversion-symmetry breaking bulk superconducting states of AM. 

\textbf{\textit{Model of a AM metal.---}}
We focus on a two-dimensional $d$-wave AM metal~\cite{vsmejkal2022emerging,vsmejkal2022beyond} with a low-energy Hamiltonian expanded near the $\Gamma$ point as
\bea
H(\bs k)=\psi_{\bs k}^\dagger \big[c_0(k_{x}^{2}+k_{y}^{2}) \sigma_0 + \lambda k_{x}k_{y}\sigma_{z} - \mu \sigma_0\big] \psi_{\bs k}
\label{eq:H_kin}
\eea
with the two-component spinor $\psi_{\bs k}^\dagger\!\equiv\!(c^\dagger_{\bs k,\uparrow}, c^\dagger_{\bs k,\downarrow})$  and the Pauli matrices act on spin space. The first term describes the kinetic energy, while $\lambda$ and $\mu$ denote the strength of AM splitting and the chemical potential, respectively. The AM breaks the time-reversal symmetry $\mathcal{T}$ and the fourfold rotational symmetry $C_4^z$ about $z$ axis, which is characterized by anisotropic spin splitting vanishing along $k_x$ and $k_y$ axes, thus forming a $d$-wave magnet. However, the system still preserves a magnetic fourfold rotational symmetry $\mathcal{C}_4^z\!\equiv\!\mathcal{T}C_4^z$ such that $\mathcal{C}_4^z H(\bs k) (\mathcal{C}_4^z)^{-1} \!=\! H(\bs k)$. As a result, the net magnetization of the system must be zero although the two bands become spin split. The band structure looks similar to those of RuO$_2$ and KRu$_4$O$_8$~\cite{vsmejkal2022beyond,gonzalez2021efficient}. Throughout this study, we fix $\lambda\!=\!c_0\!=\!100$meV.
	
There are diverse microscopic mechanism of superconducting pairing and in the following we focus on a basic effective BCS-type attractive interaction, most commonly treated for phonon-mediated pairing. The resulting effective interaction is written in momentum space as 
\bea
-U \sum_{{\bm k},{\bm k}',{\bm q}} c^{\dagger}_{{\bm k}+{\bm q}/2,\uparrow}c^{\dagger}_{-{\bm k}+{\bm q}/2,\downarrow}
c^{}_{-{\bm k}'+{\bm q}/2,\downarrow}c^{}_{{\bm k}'+{\bm q}/2,\uparrow}.
\label{eq:H_int}
\eea
Here, $U$ is the strength of onsite  interactions and the attraction holds only near the Fermi level within regions bounded by the Debye frequency $\omega_D$ as illustrated in Fig.~\ref{fig_1}. Such interactions typically lead to the condensation of spin-singlet $s$-wave $\bs q\!=\!0$ states with the order parameter $\Delta_0 \equiv \langle  c_{\bm k, \uparrow} c_{-\bm k, \downarrow} \rangle$. Alternatively, sought-after spin-singlet $s$-wave PDW states with the order parameter $\Delta_{\bs q} \equiv \langle  c_{\bm k+\bs q/2, \uparrow} c_{-\bm k+\bs q/2, \downarrow} \rangle$ carrying a finite center of mass momentum $\bs q$ can potentially be realized. Below, we focus on such $s$-wave states ignoring the long-range interactions that can induce spin-singlet $d$-wave or spin-triplet $p$-wave states~\cite{chakraborty2024zero,chakraborty2024constraints,hong2025unconventional,de2024unconventional} (see Sec.~III of the Supplementary Material for details).

\textbf{\textit{PDW in AM.---}}
The phase diagram of superconducting states can be obtained by minimizing the Ginzburg–Landau free energy. The PDW order parameter can be conveniently written as
\bea
\hat{\Delta}_{\bs q}\!=\!\{ \Delta_{\bs q} ,\Delta_{\bar{\bs q}} , \Delta_{-\bs q} ,\Delta_{-\bar{\bs q}} \},
\eea
with corresponding wavevectors $ \{ \bs q , \bar{\bs q} , -\bs q , -\bar{\bs q} \}$. Here, $\bs q$ and $\bar{\bs q}$ are connected through the magnetic fourfold rotation such that $\mathcal{C}_4^z \bs q\!=\!-\bar{\bs q}$. The free energy is constructed by imposing gauge, translational, inversion, and magnetic fourfold rotational symmetries and is written as~\cite{agterberg2008dislocations}
\bea
F&\!=\!&  \int d\bs q \biggl[ \alpha(\bs q) \sum_{i=1}^4 |\hat{\Delta}_{\bs q}^i|^2\!+\! \beta_1 \bigg(\sum_{i=1}^4  |\hat{\Delta}_{\bs q}^i|^2\bigg)^2
\nonumber
\\
&\!+\!&\beta_2 (|\Delta_{\bs q}|^2\!+\!|\Delta_{-\bs q}|^2)(|\Delta_{\bar{\bs q}}|^2\!+\!|\Delta_{-\bar{\bs q}}|^2)
\nonumber 
\\ 
&\!+\!& \beta_3 (|\Delta_{\bs q}|^2|\Delta_{-\bs q}|^2+|\Delta_{\bar{\bs q}}|^2 |\Delta_{-\bar{\bs q}}|^2)
\nonumber
\\
&\!+\!& \beta_4 (\Delta_{\bs q}\Delta_{-\bs q} \Delta_{\bar{\bs q}}^* \Delta_{-\bar{\bs q}}^* + \Delta_{\bs q}^*\Delta_{-\bs q}^*\Delta_{\bar{\bs q}} \Delta_{-\bar{\bs q}}) \biggl].
\label{eq:free}
\eea
\begin{figure}[]
\includegraphics[width = 1\columnwidth]{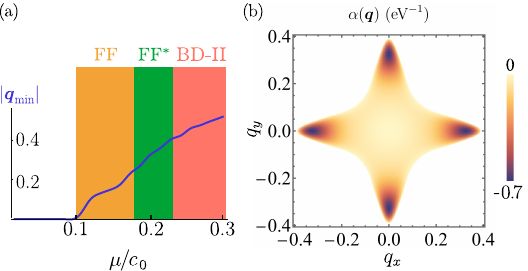}
\caption{(color online) (a) Absolute value of the ordering wavevector $|\bs q_{\text{min}}|$ as a function $\mu/c_0$. When $\mu/c_0\!\ge\!0.1$, $|\bs q_{\text{min}}|$ becomes finite which indicates the stabilization of PDW states. Symmetrically distinct PDW states are stabilized depending on the quartic coefficients of the free energy given in Eq.~(\ref{eq:free}). (b) The quadratic coefficient $\alpha(\bs q)$ for $\mu/c_0\!=\!0.2$. $\bs q_{\text{min}}$ are fourfold degenerate and lie on $q_x$ and $q_y$ axes. Here and below, we set $U\!=\!27$meV, $\omega_D\!=\!10$meV, and the temperature $T\!=\!10$K.}
\label{fig_2}
\end{figure}
The quadratic and quartic coefficients can be explicitly calculated starting from the microscopic Hamiltonian given in Eq.~(\ref{eq:H_kin}) by integrating out the electronic degrees of freedom~\cite{radzihovsky2011fluctuations,soto2014pair,liu2017unconventional,hu2019topological,sim2020multipolar,sim2022topological} (see Sec.~I of the Supplementary Material for details). Depending on the coefficients, the free energy stabilizes the $\bs q\!=\!0$ state or different types of $\bs q\neq0$ PDW states. First, the free energy stabilizes the PDW states if the minimum of $\alpha(\bs q)$ is located at $\bs q\!\neq\!0$, i.e., $\bs q_{\text{min}}\!\neq\!0$. The wavevector $\bs q_{\text{min}}$ determines the center of mass momentum that the Cooper pair carries in the ground state. Second, the system selects one of the symmetrically distinct PDW states depending on the quartic coefficients. In Table.~\ref{tab_1}, we list all possible PDW states which include inversion-breaking ones: the FF and FF* state. For these, the system may display a non-reciprocal superconducting current, since the time-reversal symmetry is broken by the AM and the inversion symmetry is spontaneously broken by the superconducting PDW. It is also worthwhile to note that the free energy contains a $U(1) \times U(1) \times U(1)$ symmetry where two additional $U(1)$ symmetries come from translation invariance along the $x$ and $y$ directions. As a result, the PDW states can induce fractional vortices except for the FF state as pointed out in Ref.~\onlinecite{agterberg2008dislocations}.
\begin{table*}[]
\begin{tabular}{|c|c|c|c|c|}
\hline
State & $\hat \Delta_{\bs q}=\{\Delta_{\bs q},\Delta_{\bar{\bs q}},\Delta_{-\bs q},\Delta_{-\bar{\bs q}}\}$& Stability~\cite{agterberg2008dislocations} & Inversion & Diode direction \\
\hline\hline
Fulde–Ferrell & $\{e^{i\phi_1},0,0,0\}$ & $\beta_2>0,\beta_2+\beta_3>0$, & X & $\bs q$ \\
(FF)&&$3\beta_2+\beta_3-|\beta_4|>0$&&\\
\hline
Fulde–Ferrell* & $\{e^{i\phi_1},e^{i\phi_2},0,0\}$& $\beta_2<0,$  & X & $\bs q, \bar{\bs q}$\\
(FF*)&&$\beta_2+\beta_3-|\beta_4|>0$&&\\
\hline\hline
Uni-directional& $\{e^{i\phi_1},0,e^{i\phi_2},0\}$& $\beta_2+\beta_3<0,$  & O & none \\
(UD) &&$\beta_2-\beta_3-|\beta_4|>0$&&\\
\hline
Bi-directional-I & $\{e^{i\phi_1},e^{i\phi_2},e^{i\phi_3},e^{i(\phi_1+\phi_3-\phi_2)}\}$&$\beta_4<0,\beta_2+|\beta_3|<|\beta_4|$,
& O & none  \\
(BD-I)&&$3\beta_2+\beta_3-|\beta_4|<0$&&\\
\hline
Bi-directional-II & $\{e^{i\phi_1},ie^{i\phi_2},e^{i\phi_3},ie^{i(\phi_1+\phi_3-\phi_2)}\}$& $\beta_4>0,\beta_2+|\beta_3|<|\beta_4|$,& O
& none \\
(BD-II)&&$3\beta_2+\beta_3-|\beta_4|<0$ &  &\\
\hline
\end{tabular}
\caption{Symmetrically distinct PDW states. The uni-directional (UD) state with the order parameter $\hat{\Delta}_{\bs q}\!=\!\{e^{i\phi_1},0,e^{i\phi_2},0 \}$ is referred to as the Larkin-Ovchinikov state in the literature. The third column quantifies the parameter regions where each state is selected. The fourth column shows whether each state spontaneously breaks the inversion symmetry. The final column specifies the directions in which the supercurrent diode effect can occur.} 
\label{tab_1}
\end{table*}
	
To investigate whether the system stabilizes PDW and which ones, we calculate the coefficients of the free energy, $\alpha(\bs q)$ and $\beta_i$. We obtain the phase diagram for varying $\mu/c_0$ as shown in Fig.~\ref{fig_2}(a). When $\mu/c_0\!<\!0.1$, the minimum of $\alpha(\bs q)$ is located at $\bs q_{\text{min}}\!=\!0$ and the system selects the $\bs q\!=\!0$ state. The system starts to stabilize the PDW state when $\mu/c_0\!\ge\!0.1$. Within such a regime, the ordering wavevector is finite, $\bs q_{\text{min}}\!\neq\!0$, and gradually increases as $\mu$ increases, reflecting the size of the normal Fermi surface as shown in Fig.~\ref{fig_2}(a). In Fig.~\ref{fig_2}(b), we plot $\alpha(\bs q)$ with $\mu/c_0\!=\!0.2$ which clearly shows that $\bs q_{\text{min}}\!\neq\!0$ and the ordering wavevectors lie on the $q_x$ and $q_y$ axes and are fourfold degenerate. The degeneracy originates from the magnetic fourfold rotational symmetry.

When $0.1\!\le\!\mu/c_0\!<\!0.18$, the quartic coefficients satisfy the following relations given in Table.~\ref{tab_1}: $\beta_2\!>\!0$, $\beta_2\!+\!\beta_3\!>\!0$, and $3\beta_2\!+\!\beta_3\!-\!|\beta_4|\!>\!0$. Within such a regime, the system selects the FF state with the order parameter $\hat{\Delta}_{\bs q}\!=\!\{e^{i\phi_1},0,0,0\}$ which spontaneously breaks the inversion symmetry. Here we  highlight that the realization of the FF state in AM does not require an external magnetic field unlike the original proposal given in Ref.~\onlinecite{fulde1964superconductivity} and Rashba spin-orbit coupled systems~\cite{agterberg2007magnetic,dimitrova2007theory,agterberg2012magnetoelectric}. At the same time, the inversion symmetry of the system is spontaneously broken by the superconducting state, unlike the Rashba systems where the inversion is broken by the spin-orbit coupling. 

For $0.18\!\le\!\mu/c_0\!<\!0.23$, the system stabilizes the FF* state with $\hat{\Delta}_{\bs q}\!=\!\{e^{i\phi_1},e^{i\phi_2},0,0\}$ which also breaks the inversion symmetry. Unlike the FF state, the order parameter of the FF* state has two phase parameters, $\phi_1$ and $\phi_2$. As a result, the system can have a half-quantum vortex as a defect though careful comparison between the energy of the conventional vortex and the half-quantum vortex will be needed~\cite{agterberg2008dislocations}. When $\mu/c_0\!\ge\!0.23$, the system selects the so-called BD-II state with $\hat{\Delta}_{\bs q}\!=\!\{e^{i\phi_1},ie^{i\phi_2},e^{i\phi_3},ie^{i(\phi_1+\phi_3-\phi_2)}\}$ which preserves the inversion symmetry. 

\textbf{\textit{Diode effect in PDW.---}}
In the following, we consider the supercurrent diode effect as a tool to distinguish the different forms of PDW states.
Near the superconducting phase transition temperature, the formation of the PDW states can be described phenomenologically by a Ginzburg–Landau functional $F \!=\! \int f (\bs r) d\bs r$, with the free energy density which can be spatially homogeneous given by
\small
\bea
\nonumber
f(\bs r) &=& \sum_{i=1}^4 \bigg[ a_0 |\hat{\Delta}_{\bs q}^i(\bs r)|^2 + a_2 |\D \hat{\Delta}_{\bs q}^i(\bs r)|^2 + a_4 |\D^2 \hat{\Delta}_{\bs q}^i(\bs r)|^2\bigg].
\\
\label{eq:free_real}
\eea
\normalsize
Here, $\D=\Grad+i(2e)\bs A$ is the covariant derivative. The supercurrent can be derived from Eq.~(\ref{eq:free_real}) by standard means~\cite{kopnin2001theory,samokhin2017current,samokhin2019fulde,yerin2023multiple}, by varying the free energy concerning the vector potential $\bs{A}$ and subsequently enforcing $\bs{A}=0$, yielding the following result (see Sec.~II of the Supplementary Material for details):
\small
\bea
\nonumber
\bs J &\equiv&-\frac{\partial f}{\partial \bs A}\bigg|_{\bs A=0}
\\
\nonumber
&=&-4ea_2 \sum_{i=1}^4 \text{Im}[\hat{\Delta}_{\bs q}^i(\bs r)^*\Grad \hat{\Delta}_{\bs q}^i(\bs r)]
\\
&-& 4ea_4 \sum_{i=1}^4 \text{Im}[(\Grad \hat{\Delta}_{\bs q}^i(\bs r))^* \Grad^2 \hat{\Delta}_{\bs q}^i(\bs r)-(\hat{\Delta}_{\bs q}^i(\bs r))^* \Grad^3 \hat{\Delta}_{\bs q}^i(\bs r)].
\nonumber \\
\label{eq:current}
\eea
\normalsize
Here, we included contributions from the fourth order derivatives as they are crucial for ensuring the stability of the $\bs q\neq0$ ground state~\cite{samokhin2017current,samokhin2019fulde}.

For the FF state with $\Delta_{\bs q}(\bs r)\!=\!|\Delta|e^{iqx}$ and $\Delta_{\bar{\bs q}}(\bs r)\!=\!\Delta_{-\bs q}(\bs r)\!=\!\Delta_{-\bar{\bs q}}(\bs r)\!=\!0$, the supercurrent is given by
\bea
J_x(q)= -4e|\Delta|^2(a_2 q+2a_4 q^3),~J_y(q) = 0.
\label{eq:current_FF}
\eea
For the FF* state with $\Delta_{\bs q}(\bs r)=|\Delta|e^{iqx}$, $\Delta_{\bar{\bs q}}(\bs r)=|\Delta|e^{iqy}$, and $\Delta_{-\bs q}(\bs r)\!=\!\Delta_{-\bar{\bs q}}(\bs r)\!=\!0$, the supercurrent is written as
\bea
J_x(q)=J_y(q)= -4e|\Delta|^2 (a_2 q+2a_4 q^3).
\label{eq:current_FF*}
\eea
For both cases, the supercurrent vanishes at $q\!=\!0$ and $q\!=\!\sqrt{-a_2/2a_4}\!=\!q_\text{min}$. We want to emphasize that the FF* state can induce a non-reciprocal supercurrent concurrently in both the $x$ and $y$ directions unlike the FF state, which can induce the non-reciprocal current only in the $x$ direction. Other PDW states, which include UD, BD-I, and BD-II states, cannot induce the non-reciprocal current, since they preserve the inversion symmetry. 
	
In equilibrium $\partial F/\partial \Delta\!=\!0$, where $F$ is the free energy given in Eq.~(\ref{eq:free}), leading to $|\Delta|^2=-\alpha(q)/(2\beta_1)$ for the FF state and $|\Delta|^2=-\alpha(q)/(4\beta_1+\beta_2)$ for the FF* state. At the same time, $\alpha(q)\!=\!a_0\!+\!a_2q^2\!+\!a_4q^4$ which does not contain linear nor cubic terms unlike the Rashba systems with external magnetic field considered in Ref.~\onlinecite{yuan2022supercurrent} and Ref.~\onlinecite{daido2022intrinsic}. Then one can rewrite the supercurrent as
\bea
J_x(q)= e \frac{\alpha(q)}{\beta_1}  \frac{\partial \alpha(q)}{\partial q},~J_y(q) = 0
\label{eq:supercurrent_FF}
\eea
for the FF state and
\bea
J_x(q)=J_y(q)= 2e \frac{\alpha(q)}{4\beta_1+\beta_2}  \frac{\partial \alpha(q)}{\partial q}
\label{eq:supercurrent_FF*}
\eea
for the FF* state. We note that the supercurrent formula in Eq.~(\ref{eq:supercurrent_FF}) has the same form as the one given in Ref.~\onlinecite{yuan2022supercurrent} for the FF state in Rashba spin-orbit coupled systems. 
\begin{figure}[]
\includegraphics[width = 1.0\columnwidth]{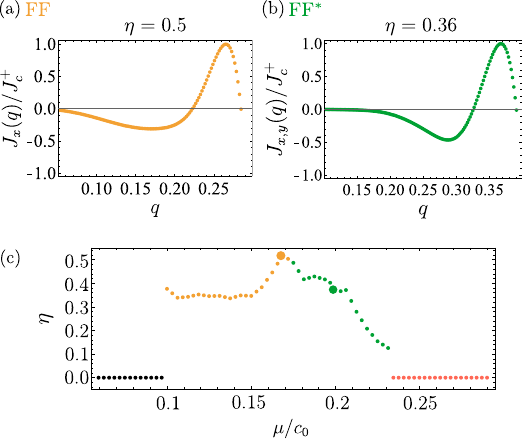}
\caption{(color online) (a) The supercurrent $J_x(q)$ as a function of momentum $q$ for FF state with $\mu/c_0\!=\!0.17$. The supercurrent diode coefficient $\eta\!=\!0.5$. (b) The supercurrent $J_x(q)$ and $J_y(q)$ as a function of $q$ for FF* state with $\mu/c_0\!=\!0.2$. The diode coefficient $\eta\!=\!0.36$. (c) The diode coefficient $\eta$ as a function of $\mu/c_0$. It reaches the maximum value of $\eta\!=\!0.5$ at $\mu/c_0\!=\!0.17$ near the phase boundary between FF state and FF* state.}
\label{fig_3}
\end{figure}
	
In Fig.~\ref{fig_3}(a), we plot the supercurrent along $x$ direction for the FF state stabilized for $\mu/c_0\!=\!0.17$, using Eq.~(\ref{eq:supercurrent_FF}). It shows the nonreciprocity of the critical current; the maximum $J_c^+$ and the minimum $-J_c^-$ are different around the $q_{\text{min}}$, where the supercurrent vanishes. Using Eq.~(\ref{eq:supercurrent_FF*}), we also plot the current relation for $\mu/c_0\!=\!0.2$, where the system stabilizes the FF* state, in Fig.~\ref{fig_3}(b). Unlike the FF state, the FF* state induces the non-reciprocal current in both the $x$ and $y$ directions. 

The non-reciprocal nature of the supercurrent can be conveniently quantified by a diode coefficient~\cite{yuan2022supercurrent} which is defined as
\bea
\eta \equiv \frac{J_c^+-J_c^-}{J_c^++J_c^-}.
\eea
In Fig.~\ref{fig_3}(c), we plot the coefficient $\eta$ as a function of $\mu/c_0$. When $\mu/c_0\!<\!0.1$, the system selects the $\bs q\!=\!0$ state preserving the inversion symmetry with vanishing $\eta$. For $0.1\!\le\!\mu/c_0\!<\!0.23$, one of the inversion breaking states, i.e., FF state and FF* state, is stabilized and $\eta$ becomes finite. When $\mu/c_0\!\ge\!0.23$, $\eta$ becomes zero again, since the system stabilizes the BD-II state and preserves the inversion symmetry. Overall, the supercurrent diode effect can be utilized to detect different types of PDW states in future experiments.

\textbf{\textit{Summary and outlook.---}}
We have studied bulk superconductivity in a two-dimensional metallic system with $d$-wave AM spin splitting in momentum space. We focused  on the emergence of superconductivity driven by onsite attractive interactions analyzed through a microscopically derived Ginzburg–Landau theory. We find that symmetrically distinct PDW states can be stabilized as a function of varying chemical potential $\mu$, including the inversion-breaking FF and FF* states. We then investigated the supercurrent properties and found a non-zero non-reciprocity; i.e. the FF state induces a non-reciprocal supercurrent in one direction, while the FF* state can induce it in two orthogonal directions.
	
There have been many theoretical proposals for realizing superconductivity of different types in AM~\cite{chakraborty2024zero,bose2024altermagnetism}. In the future it would worthwhile to explore PDW formation with different pairing symmetries and potential topological properties, for example with spin-triplet pairing or in the presence of Rashba spin-orbit coupling~\cite{PhysRevB.108.184505}. Beyond collinear AM other real space non-collinear spin textures might show similar effects~\cite{lee2024fermi}.

Recent research has identified several promising material candidates for studying the interplay between altermagnetism and superconductivity. RuO$_2$ thin films, which exhibit d-wave altermagnetic characteristics with momentum-dependent spin splitting and zero net magnetization, demonstrate superconductivity under strain~\cite{uchida2020superconductivity,ruf2021strain}. Hole-doped La$_2$CuO$_4$, known for high-temperature superconductivity, offers a rich context for exploring PDW states and the supercurrent diode effect~\cite{agterberg2020physics}. While other metallic altermagnets, such as KRu4O8~\cite{vsmejkal2022beyond}, MnTe~\cite{krempasky2024altermagnetic}, and CrSb~\cite{reimers2024direct}, have not yet shown superconductivity, they could be tuned via pressure or proximity effects. These materials, supported by microscopic theoretical modelling, provide exciting opportunities for experimental realisation of non-reciprocal effects of superconductivity.

\textbf{\textit{Acknowledgements.---}} 
We thank P. Rao, J. Habel, M. J. Park, V. Leeb, R.-X. Zhang, and L. {\v{S}}mejkal for insightful discussions related to this work. We acknoweldge M. Scheurer, I. Eremin, R. Fernandes, and D. Agterberg for comments on the manuscript. We especially thank D. Agterberg for pointing out an earlier flaw in the derivation of Eq.~(\ref{eq:current_FF}). G.B.S. is funded by the European Research Council (ERC) under the European Unions Horizon 2020 research and innovation program (grant agreement No. 771537). J. K. acknowledges support from the Imperial-TUM flagship partnership. The research is part of the Munich Quantum Valley, which is supported by the Bavarian state government with funds from the Hightech Agenda Bayern Plus. G.B.S was supported by Basic Science Research Program through the National Research Foundation of Korea (NRF) (RS-2024-00453943).

\renewcommand{\thefigure}{S\arabic{figure}}
\setcounter{figure}{0}
\renewcommand{\theequation}{S\arabic{equation}}
\setcounter{equation}{0}

\begin{widetext}
\section{Supplementary Material}

\subsection{I. Ginzburg-Landau free energy and one-loop expansion}
In this section, we compute the quadratic and quartic coefficients in the Ginzburg-Landau free energy and see how the symmetrically distinct pair density waves (PDW) states are stabilized.  We first introduce the free electron and hole propagator
\bea
\nonumber
G_e(\omega,\bs k) &=& \frac{1}{i\omega - H(\bs k )},\\
G_h(\omega,\bs k) &=& \frac{1}{i\omega + H^T(- \bs k )}
\eea
where $\omega=2\pi (n+{{1}\over{2}})T$ is the Matsubara frequency with $T$ being the temperature. The non-interacting Hamiltonian given in the main text is written as
\bea
H(\bs k)=c_0(k_{x}^{2}+k_{y}^{2}) \sigma_0 + \lambda k_{x}k_{y}\sigma_{z} - \mu \sigma_0
\eea
where $\lambda$ and $\mu$ denote the strength of the altermagnetism and the chemical potential respectively. The Ginzburg-Landau free energy for the PDW states in our system, which respects gauge, translational, inversion, and magnetic fourfold rotational symmetries, is written as
\bea
F=  \int d\bs q f(\bs q)
\eea
where
\bea
f (\bs q)=\biggl[ \alpha(\bs q) \sum_{i=1}^4 |\hat{\Delta}_{\bs q}^i|^2+ \beta_1 \bigg(\sum_{i=1}^4  |\hat{\Delta}_{\bs q}^i|^2\bigg)^2 + \beta_2 (|\Delta_{\bs q}|^2\!+\!|\Delta_{-\bs q}|^2)(|\Delta_{\bar{\bs q}}|^2\!+\!|\Delta_{-\bar{\bs q}}|^2)
\nonumber 
\\ 
+ \beta_3 (|\Delta_{\bs q}|^2|\Delta_{-\bs q}|^2+|\Delta_{\bar{\bs q}}|^2 |\Delta_{-\bar{\bs q}}|^2) + \beta_4 (\Delta_{\bs q}\Delta_{-\bs q} \Delta_{\bar{\bs q}}^* \Delta_{-\bar{\bs q}}^* + \Delta_{\bs q}^*\Delta_{-\bs q}^*\Delta_{\bar{\bs q}} \Delta_{-\bar{\bs q}}) \biggl].
\eea
Here, The PDW order parameter is written as $\hat{\Delta}_{\bs q}\!=\!\{ \Delta_{\bs q} ,\Delta_{\bar{\bs q}} , \Delta_{-\bs q} ,\Delta_{-\bar{\bs q}} \}$
with the corresponding wavevectors $ \{ \bs q , \bar{\bs q} , -\bs q , -\bar{\bs q} \}$. Ref.~\onlinecite{agterberg2008dislocations,fradkin2015colloquium,agterberg2020physics} provided an identical free energy for a system with the fourfold rotational symmetry of a square lattice. For later simplicity, we rewrite it as
\bea
f(\bs q)= \alpha(\bs q) \sum_{i=1}^4 |\hat{\Delta}_{\bs q}^i|^2+ b_1 \sum_{i=1}^4  |\hat{\Delta}_{\bs q}^i|^4 + b_2 (|\Delta_{\bs q}|^2\!+\!|\Delta_{-\bs q}|^2)(|\Delta_{\bar{\bs q}}|^2\!+\!|\Delta_{-\bar{\bs q}}|^2)
\nonumber 
\\ 
+ b_3 (|\Delta_{\bs q}|^2|\Delta_{-\bs q}|^2+|\Delta_{\bar{\bs q}}|^2 |\Delta_{-\bar{\bs q}}|^2) + b_4 (\Delta_{\bs q}\Delta_{-\bs q} \Delta_{\bar{\bs q}}^* \Delta_{-\bar{\bs q}}^* + \Delta_{\bs q}^*\Delta_{-\bs q}^*\Delta_{\bar{\bs q}} \Delta_{-\bar{\bs q}}).
\label{eq:s_free}
\eea
The quadratic coefficient in Eq.~(\ref{eq:s_free}) is given as~\cite{radzihovsky2011fluctuations,soto2014pair}
\bea
\alpha(\bs q)&=&\frac{1}{U}-\frac{T}{2}\sum_{\omega,\bs k} \text{tr} [G_e(\omega,\bs k+\bs q/2) \sigma_y G_h(\omega,\bs k+\bs q/2) \sigma_y^\dagger].
\label{eq:s_quadratic}
\eea
Here, the sum over $\bs k$ is done near the Fermi level within regions bounded by the Debye frequency $\omega_D$. In this work, we set $c_0=100$meV, $\omega_D=10$meV, and $T=10$K. We plot $\alpha(\bs q)$ for $\lambda/c_0=0.3$, $\mu/c_0=0.15$, and $U=15$meV using Eq.~(\ref{eq:s_quadratic}) in Fig.~\ref{fig_s1}(a). The minimum is located at $\bs q=0$. In Fig.~\ref{fig_s1}(b), we plot $\alpha(\bs q)$ for $\lambda/c_0=0.9$, $\mu/c_0=0.15$, and $U=22$meV. The minimum is located at $\bs q \neq 0$ and the system selects the PDW state, which carries the finite center of mass momentum. We also obtain the phase diagram as shown in Fig.~\ref{fig_s1}(c), which indicates the emergence of PDW states in the presence of band splitting caused by the altermagnetism. In Fig.~\ref{fig_s1}(d), we plot $|\frac{T}{2}\sum_{\omega,\bs k} \text{tr} [G_e(\omega,\bs k+\bs q/2) \sigma_y G_h(\omega,\bs k+\bs q/2) \sigma_y^\dagger]|$, absolute value of the second term in Eq.~(\ref{eq:s_quadratic}). It corresponds to minimum strength of attractive interaction, $U_\text{min}$, required to stabilize the superconducting state.
\begin{figure}[]
	\includegraphics[width = 0.8\columnwidth]{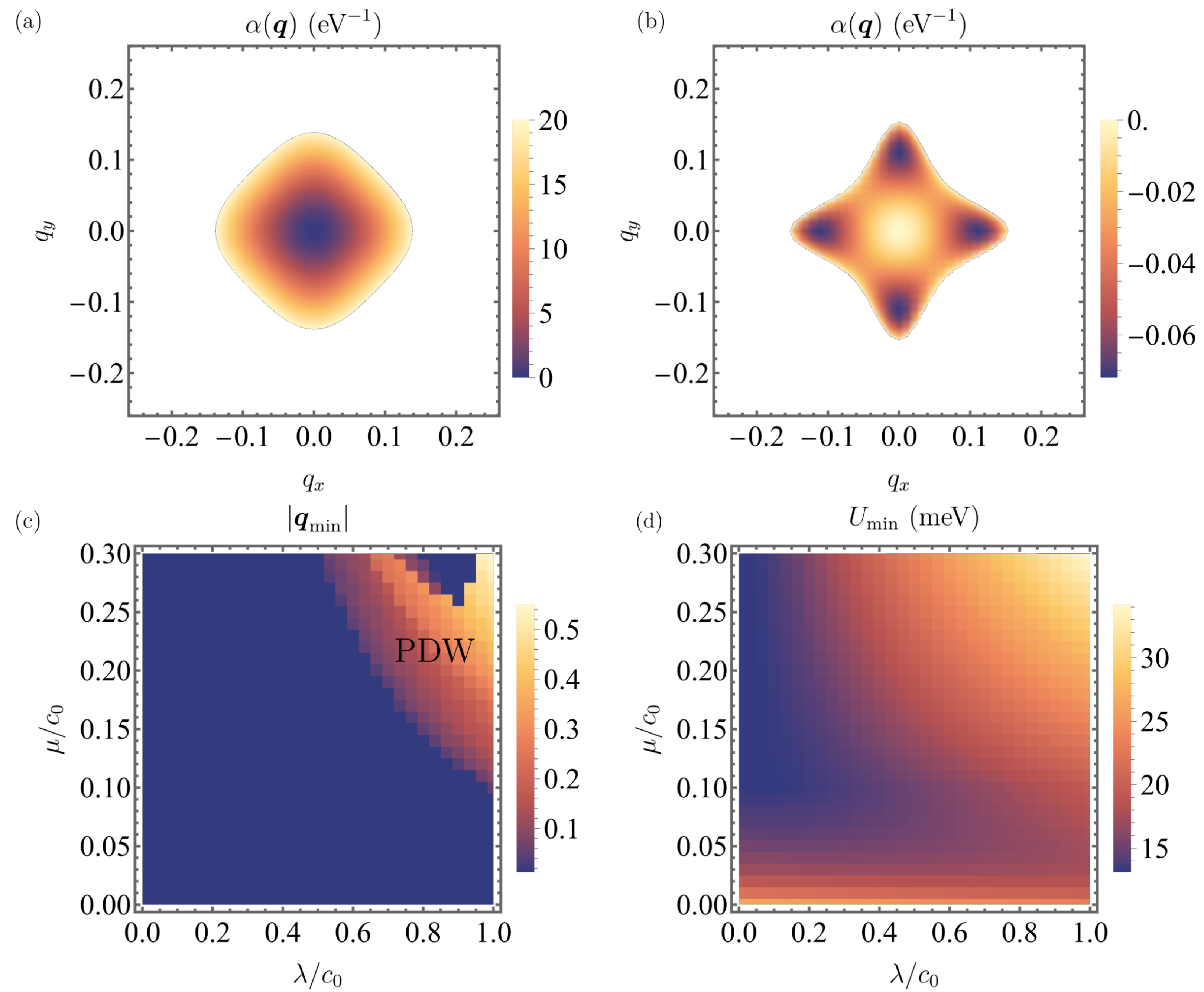}		
	\caption{(color online) (a) The quadratic coefficient $\alpha(\bs q)$ of the free energy for  $\lambda/c_0=0.3$, $\mu/c_0=0.15$, and $U=15$meV. (b) $\alpha(\bs q)$ for $\lambda/c_0=0.9$, $\mu/c_0=0.15$, and $U=22$meV. $\bs q_{\text{min}}$ are fourfold degenerate and lie on $q_x$ and $q_y$ axes. The degeneracy originates from the magnetic fourfold rotational symmetry of the system. (c) $|\bs q_{\text{min}}|$ as functions $\lambda/c_0$ and $\mu/c_0$. In the regime where $|\bs q_{\text{min}}|\neq0$, the PDW state is stabilized. (d) Minimum strength of attractive interaction $U_\text{min}$ required to stabilize the superconducting state.}
	\label{fig_s1}
\end{figure}
The quartic coefficients in Eq.~(\ref{eq:s_free}) are written as
\small
\bea
\nonumber
b_1&=&T\sum_{\omega,\bs k} \text{tr} [ G_e(\omega,\bs k+ \frac{\bs q_\text{min}}{2}) \sigma_y G_h (\omega,-\bs k+\frac{\bs q_\text{min}}{2}) \sigma_y^\dagger  G_e(\omega,\bs k+\frac{\bs q_\text{min}}{2}) \sigma_y G_h (\omega,-\bs k+\frac{\bs q_\text{min}}{2}) \sigma_y^\dagger],
\\ \nonumber
b_2&=& 4T \sum_{\omega,\bs k} \text{tr} [ G_e(\omega,\bs k+\frac{\bs q_\text{min} +\bar{\bs q}_\text{min}}{2}) \sigma_y G_h (\omega,-\bs k+\frac{\bs q_\text{min}-\bar{\bs q}_\text{min}}{2}) \sigma_y^\dagger  G_e(\omega,\bs k+\frac{\bs q_\text{min}+\bar{\bs q}_\text{min}}{2}) \sigma_y G_h (\omega,-\bs k+\frac{-\bs q_\text{min} + \bar{\bs q}_\text{min}}{2}) \sigma_y^\dagger],
\\ \nonumber
b_3&=& 4T \bigg[ \sum_{\omega,\bs k} \text{tr} [ G_e(\omega,\bs k+\frac{\bs q_\text{min}}{2}) \sigma_y G_h (\omega,-\bs k+\frac{\bs q_\text{min}}{2}) \sigma_y^\dagger  G_e(\omega,\bs k+\frac{\bs q_\text{min}}{2}) \sigma_y G_h (\omega,-\bs k - \frac{3 \bs q_\text{min}}{2}) \sigma_y^\dagger]
\\ \nonumber
&& ~~~+\sum_{\omega,\bs k} \text{tr} [ G_e(\omega,\bs k-\frac{\bs q_\text{min}}{2}) \sigma_y G_h (\omega,-\bs k-\frac{\bs q_\text{min}}{2}) \sigma_y^\dagger  G_e(\omega,\bs k-\frac{\bs q_\text{min}}{2}) \sigma_y G_h (\omega,-\bs k + \frac{3 \bs q_\text{min}}{2}) \sigma_y^\dagger] \bigg],
\\ \nonumber
b_4&=& 4T \sum_{\omega,\bs k} \text{tr} [ G_e(\omega,\bs k+\frac{\bs q_\text{min} + \bar{\bs q}_\text{min}}{2}) \sigma_y G_h (\omega,-\bs k+\frac{\bs q_\text{min}-\bar{\bs q}_\text{min}}{2}) \sigma_y^\dagger  G_e(\omega,\bs k+\frac{-\bs q_\text{min}- \bar{\bs q}_\text{min}}{2}) \sigma_y G_h (\omega,-\bs k+\frac{-\bs q_\text{min} + \bar{\bs q}_\text{min}}{2}) \sigma_y^\dagger]\\
\label{eq:s_quartic}
\eea
\normalsize
where the symmetry factor 4 is coming from the cyclic permutation of the trace. Utilizing Eq.~(\ref{eq:s_quartic}), we plot the quartic coefficients in Fig.~\ref{fig_s2}(a) for $\lambda/c_0=1$ while varying $\mu/c_0$, corresponding precisely to the parameter regime outlined in the main text. Depending on the value of the coefficients, the system stabilizes symmetrically distinct PDW states.
\begin{figure}[h!]
	\includegraphics[width = 0.5\columnwidth]{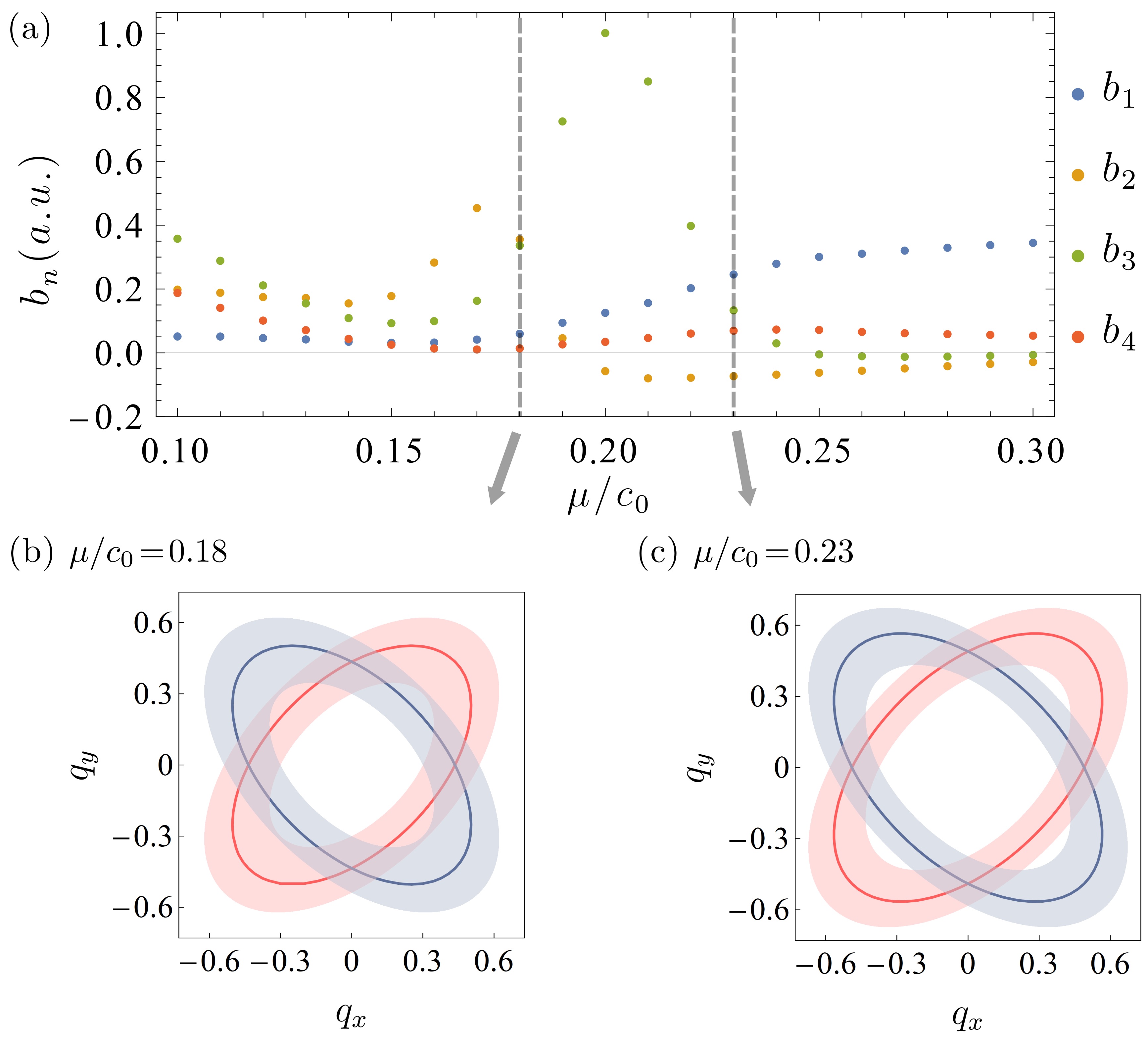}		
	\caption{(color online) (a) The quartic coefficients of the free energy as a function of $\mu/c_0$ with $\lambda/c_0=1$. When $\mu/c_0$ is less than $0.18$, the system stabilizes the Fulde-Ferrell state. In the range of $0.18$ to $0.23$, the Fulde-Ferrell* state is favored, and as $\mu/c_0$ further increases beyond $0.23$, the system stabilizes the Bi-directional-II state. (b,c) Red and blue ellipses represent the spin-polarized Fermi surfaces for $\mu/c_0=0.18$ and $0.23$. The transparent areas surrounding these ellipses, bounded by the Debye frequency $\omega_D$, indicate the regions where the pairing can happen.}
	\label{fig_s2}
\end{figure}

\subsection{II. The diode effect in inversion breaking states}
Here we formulate the supercurrent in PDW states starting from the Ginzburg-Landau functional. The free energy density in real space is written as
\bea
f(\bs A,\bs r)\!=\!\sum_{n=\bs q, \bar{\bs q}, -\bs q, -\bar{\bs q}}\bigg[ a_0 |\Delta_n(\bs r)|^2 + a_2 |\D \Delta_n(\bs r)|^2 + a_4 |\D^2 \Delta_n(\bs r)|^2 \bigg] + \frac{\bs B^2}{8\pi}
\label{eq:s_free_real}
\eea
where $\D=\Grad+i(2e)\bs A$ is the covariant derivative and $\bs B$ is the external magnetic field. 
To minimize $f(\bs A, \bs r)$ with respect to $\bs A$, we make the substitution $\bs A \to \bs A + \bs a$ in Eq.~(\ref{eq:s_free_real}) and collect the linear terms in $\bs a$:
\small
\bea
\nonumber
f(\bs A + \bs a,\bs r)&=&f(\bs A,\bs r) + a_2 \sum_{n} \bigg[  \big[\Grad \Delta_n(\bs r) \big]^* \cdot \big[ 2 i e \bs a \Delta_n(\bs r)\big] + \big[2ie \bs a \Delta_n(\bs r) \big]^* \cdot \big[ \Grad \Delta_n(\bs r)\big] + \mathcal{O}(\bs A) \bigg] 
\nonumber \\
&&+a_4 \sum_{n} \bigg[  \big[\Grad^2 \Delta_n(\bs r) \big]^* \big[ 2 i e \bs a \cdot (\Grad \Delta_n(\bs r))\big] +\big[ 2 i e \bs a \cdot (\Grad \Delta_n(\bs r))\big]^* \big[\Grad^2 \Delta_n(\bs r) \big] + \mathcal{O}(\bs A^3) \bigg]
\nonumber \\
&&+a_4 \sum_{n} \bigg[ \big[\Grad^2 \Delta_n(\bs r) \big]^* \big[ 2 i e \Grad \cdot (\bs a \Delta_n(\bs r))\big] + \big[ 2 i e \Grad \cdot (\bs a \Delta_n(\bs r))\big]^* \big[\Grad^2 \Delta_n(\bs r) \big] + \mathcal{O}(\bs A^3) \bigg] + \frac{1}{4\pi} (\Grad \times \bs A) \cdot (\Grad \times \bs a)
\nonumber \\
\label{eq:s_free_1}
\\
&=& f(\bs A,\bs r) + a_2 \sum_{n}\bigg[ 2ie\big(\big[\Grad \Delta_n(\bs r)\big]^*\Delta_n(\bs r) - \Delta_n(\bs r)^*\big[\Grad \Delta_n(\bs r)\big]\big) \cdot \bs a + \mathcal{O}(\bs A) \bigg] 
\nonumber \\
&&+ a_4 \sum_{n} \bigg[ 2ie \big( \big[\Grad^2 \Delta_n(\bs r) \big]^* \big[\Grad \Delta_n(\bs r)\big] - \big[\Grad \Delta_n(\bs r)\big]^* \big[\Grad^2 \Delta_n(\bs r) \big]\big) \cdot \bs a + \mathcal{O}(\bs A^3) \bigg]
\nonumber \\
&&- a_4 \sum_{n} \bigg[ 2ie \big( \big[\Grad^3 \Delta_n(\bs r) \big]^* \big[ \Delta_n(\bs r)\big] - \big[\Delta_n(\bs r)\big]^* \big[\Grad^3 \Delta_n(\bs r) \big]\big) \cdot \bs a + \mathcal{O}(\bs A^3) \bigg]+ \frac{1}{4\pi} (\Grad \times \bs B) \cdot \bs a
\label{eq:s_free_2}
\eea
\normalsize
where we have used $\nabla \times \bs{A} = \bs{B}$ in Eq.~(\ref{eq:s_free_1}) and Gauss's theorem in Eq.~(\ref{eq:s_free_2}). Here, $\mathcal{O}(\bs{A})$ includes terms up to first order in $\bs{A}$, and $\mathcal{O}(\bs{A}^3)$ includes terms up to third order in $\bs{A}$. Below, we use the London gauge, where $\bs{A} \to 0$ in the interior of the bulk superconductor, and discard such terms. At the minimum, the coefficient of the linear terms in $\bs{a}$ must vanish. Then we obtain
\bea
\nonumber
4ea_2 \sum_{n} \text{Im}[\Delta_n(\bs r)^*\Grad \Delta_n(\bs r)] + 4ea_4 \sum_{n} \text{Im} \big[ (\Grad \Delta_n(\bs r))^* \Grad^2 \Delta_n(\bs r) - (\Delta_n(\bs r))^* \Grad^3 \Delta_n(\bs r) \big] + \frac{1}{4\pi} (\Grad \times \bs B)= 0.
\\
\eea
With the Ampere's law we find
\bea
\nonumber
\bs J = \frac{1}{4\pi} (\Grad \times \bs B) = -4ea_2 \sum_{n} \text{Im}[\Delta_n(\bs r)^*\Grad \Delta_n(\bs r)] - 4ea_4 \sum_{n} \text{Im} \big[(\Grad \Delta_n(\bs r))^* \Grad^2 \Delta_n(\bs r) - (\Delta_n(\bs r))^* \Grad^3 \Delta_n(\bs r)\big].
\\
\label{eq:s_current}
\eea

\subsection{III. Leading pairing instability: \texorpdfstring{$s$}{s}-wave versus \texorpdfstring{$d$}{d}-wave}
\label{sec:sd_competition}

In the main text, we focus on the case where $s$-wave pairing dominates over other symmetry channels. In this section, we investigate whether this dominance persists when the $d$-wave component of the attractive interaction is present, and explore how the leading pairing symmetry changes depending on the relative interaction strengths.

We consider a general spin-singlet pairing interaction of the form
\begin{equation}
	H_p = -\sum_{\mathbf{k},\mathbf{k}',\mathbf{q}} 
	\sum_{l} U_l \, \gamma_l(\hat{\mathbf{k}})\gamma_l(\hat{\mathbf{k}}') \,
	c^{\dagger}_{\mathbf{k}+\mathbf{q}/2,\uparrow}
	c^{\dagger}_{-\mathbf{k}+\mathbf{q}/2,\downarrow}
	c^{\phantom{\dagger}}_{-\mathbf{k}'+\mathbf{q}/2,\downarrow}
	c^{\phantom{\dagger}}_{\mathbf{k}'+\mathbf{q}/2,\uparrow},
	\label{eq:interaction_sd}
\end{equation}
where $U_l$ denotes the attractive interaction strength in the pairing channel labeled by $l$, and $\gamma_l(\hat{\mathbf{k}})$ is the corresponding form factor. We focus on the spin-singlet $s$-wave and $d$-wave channels, with form factors given by
\begin{align}
	\gamma_s(\hat{\mathbf{k}}) &= 1, \\
	\gamma_{d_{x^2 - y^2}}(\hat{\mathbf{k}}) &= \sqrt{2}(\hat{k}_x^2 - \hat{k}_y^2), \\
	\gamma_{d_{xy}}(\hat{\mathbf{k}}) &= 2\sqrt{2} \hat{k}_x \hat{k}_y.
\end{align}
Below, we set $U_d \equiv U_{d_{x^2 - y^2}} = U_{d_{xy}}$ for simplicity. The order parameter in each pairing channel is defined as the expectation value of the Cooper pair field:
\begin{equation}
	\Delta_l(\mathbf{q}) \equiv \langle \gamma_l(\hat{\mathbf{k}}) c_{\mathbf{k}+\mathbf{q}/2 ,\uparrow} c_{-\mathbf{k}+\mathbf{q}/2,\downarrow} \rangle,
\end{equation}
where $\mathbf{q}$ is the center-of-mass momentum of the pair, and $\Delta_l(\mathbf{q})$ is the complex amplitude of the spin-singlet PDW state in the $l$-wave channel.

To determine the leading pairing instability, we follow the one-loop expansion method described in Section~I of the Supplemental Material. In particular, the quadratic coefficient $\alpha(\mathbf{q})$ for each pairing channel is computed using
\begin{equation}
	\alpha_l(\mathbf{q}) = \frac{1}{U_l} - \frac{T}{2} \sum_{\omega, \mathbf{k}} \gamma_l^2(\hat{\mathbf{k}})
	\mathrm{tr}\left[ G_e(\omega, \mathbf{k} + \mathbf{q}/2) \, \sigma_y \, G_h(\omega, \mathbf{k} + \mathbf{q}/2) \, \sigma_y^\dagger \right],
\end{equation}
where the Matsubara frequency sum and momentum integral are performed near the Fermi surface within an energy window set by the Debye cutoff $\omega_D$. For each point in the $(\lambda/c_0, \mu/c_0)$ plane, we evaluate $\alpha_s(\mathbf{q})$ and $\alpha_d(\mathbf{q})$ and identify the pairing channel with the lowest value of $\alpha_l(\mathbf{q})$ at its global minimum $\mathbf{q}_{\text{min}}$. Figure~\ref{fig:sd_compete}(a) shows the result for $U_s = 40$\,meV and $U_d = 30$\,meV. In this case, the $s$-wave channel remains dominant across the entire parameter range. This regime, where $U_s \gg U_d$, corresponds to the situation considered in the main text. In contrast, when $U_d$ is increased to $35$\,meV, as shown in Fig.~\ref{fig:sd_compete}(b), the $d$-wave channel becomes the leading instability in a broad region of the phase diagram.
\begin{figure}[t]
	\centering
	\includegraphics[width=0.75\columnwidth]{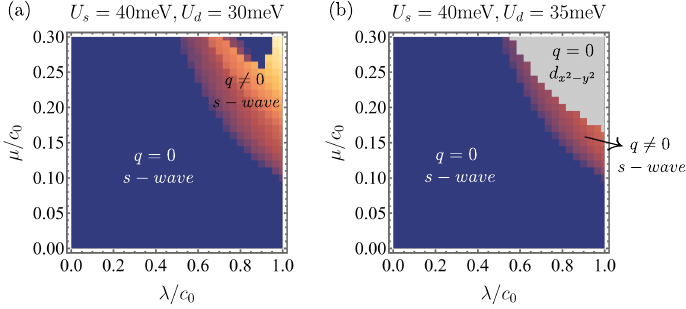}
	\caption{(color online) Leading pairing symmetry in the $(\lambda/c_0,\mu/c_0)$ plane.
		(a) $U_s = 40$\,meV, $U_d = 30$\,meV: $s$-wave pairing dominates.
		This corresponds to the parameter regime considered in the main text.
		(b) $U_s = 40$\,meV, $U_d = 35$\,meV: $d$-wave pairing becomes dominant over a sizable region.
		The parameters used are $c_0 = 100$\,meV, $\omega_D = 10$\,meV, and $T = 10$\,K, consistent with the main text and previous supplemental sections.}
	\label{fig:sd_compete}
\end{figure}

These results show that the leading pairing symmetry depends sensitively on the relative strength of the attractive interactions in the $s$-wave and $d$-wave channels. Even within the spin-singlet sector, sufficiently strong anisotropic attraction can favor a $d$-wave order parameter over the $s$-wave state as proposed in Refs.~\onlinecite{chakraborty2024zero,hong2025unconventional}.

\end{widetext}

\bibliography{alter_pdw_bib.bib}
	
\end{document}